\documentclass[a4paper,11pt]{article}
\pdfoutput=1 

\usepackage{jinstpub} 

\title{\boldmath Optimization and first tests of the experimental setup to investigate the double-polarized DD-fusion reactions}

\author[a, 1]{A.~Solovev,\note{Corresponding author.}}
\author[a]{A.~Andreyanov,}
\author[b]{L.~Barion,}
\author[b]{G.~Ciullo,}
\author[c]{R.~Engels,}
\author[a]{V.~Fotyev,}
\author[a]{K.~Ivshin,}
\author[a]{L.~Kochenda,}
\author[a]{P.~Kravchenko,}
\author[a]{P.~Kravtsov,}
\author[a]{V.~Larionov,}
\author[a]{A.~Rozhdestvensky,}
\author[a]{S.~Sherman,}
\author[a]{I.~Solovyev,}
\author[a]{V.~Trofimov,}
\author[a]{A.~Vasilyev,}
\author[a]{and M.~Vznuzdaev}

\affiliation[a]{Petersburg Nuclear Physics Institute named by B.P. Konstantinov of National Research Center "Kurchatov Institute"\\188300 Gatchina, mkr. Orlova Roshcha 1, Russian Federation}
\affiliation[b]{Dipartimento di Fisica e Scienze della Terra, Universit\`a degli Studi di Ferrara and Istituto Nazionale di Fisica Nucleare\\I-44122 Ferrara, Italy}
\affiliation[c]{Institut f\"ur Kernphysik (IKP), Forschungszentrum J\"ulich (FZJ)\\52428 J\"ulich, Germany}

\emailAdd{solovev\_an@pnpi.nrcki.ru}

\abstract{The study of DD reactions, especially with polarized reactants, helps for better understanding of the processes taking place in nuclear astrophysics and fusion reactors. At PNPI Gatchina, Russia, the PolFusion experiment with crossing of two polarized beams, i.e. a deuteron and a deuterium beam, is able to measure angular distributions of the differential cross section and, therefore, the spin-correlations coefficients with different combinations of the adjustable nuclear polarization of both beams with a center-of-mass energy between 10 to 100~keV. Some improvements and fine-tuning of the polarized ion source are performed and presented. The atomic beam source for the jet target has been modified as well. An unpolarized experiment with a 10~keV ion beam and heavy water vapor as a target has been carried out with successful registration of the fusion products.}


\collaboration[c]{of the PolFusion collaboration}

\proceeding{The International Conference "Instrumentation for Colliding Beam Physics" (INSTR20)\\
 24 - 28 February, 2020\\
  Budker Institute of Nuclear Physics, and Novosibirsk State University, Novosibirsk, Russia}

\begin{document}
\maketitle
\flushbottom

\section{Introduction}
\label{sec:intro}

The nuclear reactions with deuterons as reactants are of great importance in both, nuclear astrophysics and fusion energy applications. Studing these reactions allows to infer implications to other fields such as elementary particle physics. The reactions $^{2}H(d,p)^{3}H$ and $^{2}H(d,n)^{3}He$ are involved in primordial nucleosynthesis, specifically in the production of the $^{2}H$, $^{3}He$, and $^{4}He$ during the early universe. The energy range of the processes extends from a few to hundreds of keV. Besides primordial nucleosynthesis the deuterium is burned during the earliest stage of star evolution in the pre-main sequence phase.

These reactions together with tritium-induced reactions are the suitable processes used to fuel magnetic-~\cite{a} and inertial-confinement~\cite{b} fusion reactors to provide sufficient energy for commercial use. These reactors are expected to operate in the temperature range of kT = 1 to 30~keV. It exists some previous proposals~\cite{a} of reactors without an external tritium source due to the lack of worldwide tritium production. Initially, the reactor would operate on DD reactions gradually producing tritium in the plasma through the neutron channel of the reactions and in a special $^{6}LiD$ blankets through the known reaction $^{6}Li+n \rightarrow t+\alpha$. Afterwards, the tritium diffuses back into the reactor plasma. The reactor would eventually enter into the normal mode when accumulating a sufficient amount of tritium to initiate the DT reaction.

In addition, the DD reactions become more attractive when polarized deuterons are used. For the $^{3}H(d,n)^{4}He$ reaction ref.~\cite{c} shows the possibility to increase the cross section of the reactions in a fusion reactor by 50\% while using fuel particles with parallel spins. Another possible advantage~\cite{c} is the reduction of the neutron-producing channel branching ratio in favor of the charged-particle channel and the handling of the emission direction of reaction products. The enhancements would lower the requirements needed to reach ignition at the fusion reactors and extend the lifetime of the reactor by diminishing neutron degradation of critical components.

\section{PolFusion Experiment}
\label{sec:polfu}

Several analyzing power experiments~\cite{d,e,f} investigating the $^{2}H(d,p)^{3}H$ and $^{2}H(d,n)^{3}He$ reactions were performed in the past years with a polarized beam of deuterons bombarding an unpolarized deuterium target. The first proposal by the Kurchatov Institute group, Russia, on the experimental study of the polarization coefficients in double-polarized DD collisions dates back to 1976~\cite{t}. However, due to serious difficulties none spin-correlation experiment that includes a polarized beam and polarized target was carried out yet. 

The main problem is the low cross-sections of the reactions in the low-energy region due to the Coulomb barrier. It leads to very low count rates that requires long-term stable and automated measurements. The background from cosmic rays, terrestrial radioactivity, and from electronic disturbances are a matter of concern.

An polarized solid deuterium target is very difficult to produce and impossible to use in $4 \pi$ detector geometry. In addition, the electron screening of the nuclear charges under laboratory conditions differs from the one in the fusion reactor plasma due to the different environment of the deuterons~\cite{g}. 

PNPI Gatchina, Russia, in collaboration with Forschungszentrum J\"ulich, Germany, and INFN / University of Ferrara, Italy, has presented world's first low-energy colliding-beam experiment PolFusion~\cite{h} with polarized beam and target. 

In the PolFusion experiment, the spin-dependence of the differential cross section is measured for different spin settings of the beam and target to deduce the vector and tensor analysing powers and the spin correlation coefficients of the $^{2}H(d,p)^{3}H$ and $^{2}H(d,n)^{3}He$ reactions. From these values, the quintet suppression factor can be deduced. The results will allow to solve the discrepancy between different theoretical predictions.

\section{Polarized Sources}
\label{sec:sources}

The main idea of the experiment is to cross a polarized deuteron beam with an energy range of 10-100~keV with a polarized atomic deuterium beam having a thermal energy of about 0.01~eV. Therefore, the experimental setup consists of two polarized sources, a detector system, and polarimetry.

\subsection{Polarized Ion Source (POLIS)}
\label{sec:POLIS}

The ion source POLIS~\cite{i}, donated by the Kernfysisch Versneller Instituut (KVI) of Rijksuniversiteit Groningen to PNPI Gatchina, has undergone significant changes throughout past years while being in Gatchina. Previously, it was used as an injector with a beam energy up to 35 keV to feed a cyclotron. To fulfill the PolFusion experiment demands, the ion source should operate with a beam energy up to 100 keV and a long-term stable beam intensity.

First, some slight modifications of the dissociator were made to improve its operation condition and increase the atomic beam flow. The atomic beam intensity depends on a number of parameters: inlet gas flow, RF power, nozzle temperature, intra-beam and background scattering processes. 

In~the~dissociator, deuterium molecules are dissociated into atoms by a radio-frequency (RF) induced discharge. The RF power of 250~W at a frequency of 13.6~MHz is induced by magnetic coupling with a coil around the borosilicate glass vessel. In addition, the~RF is capacitively coupled to the plasma. The discharge area was increased by changing the coil parameters. To drastically diminish reflected RF power, a suitable matching circuit has been developed and successfully used in the RF circuit. The reflected RF power has been decreased from 15\% to about 5\% of the forward power, i.e., down to 13 W. In addition, the new matching circuit allows to avoid the necessity of fine-tuning of the RF circuit every time the dissociator was turned on.

\begin{figure}[htbp]
\centering 
\includegraphics[width=.8\textwidth,origin=c,angle=0]{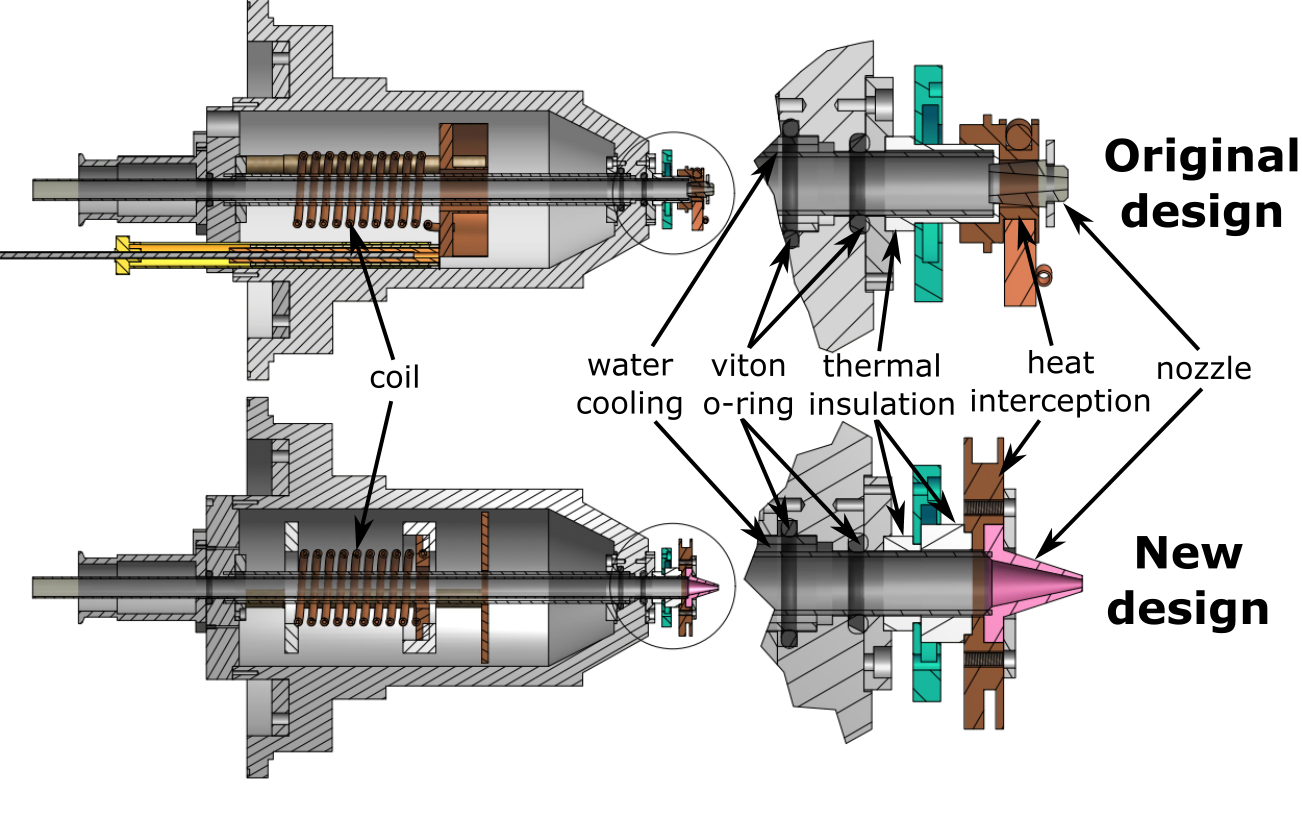}
\caption{\label{fig:i} Comparison of the original and the new nozzle design.}
\end{figure}

The $D_{2}$ gas with known flux enters the discharge tube from one end, and $D$ atoms stream out at the other. The atoms are formed into a beam by a nozzle made from aluminum alloy with a tapered canal 20~mm long and an opening 1.3~mm width. The nozzle system was modified in such a way that no gas pockets exist and no material was allowed in the plasma area that could be burned by the hot plasma and subsequently contaminate the inner nozzle surface. A comparison of the original design with the current design is presented in Figure~\ref{fig:i}. The nozzle is cooled to 65~K by a cryogenic refrigerator. The aluminum alloy has been used as nozzle material because of its high heat conductivity, especially in a cryogenic range, good mechanical properties and also the aluminium oxide surface which is better than copper for thin ice layer formation. There is a good evidence that the nozzle covered with ice has a reduced surface recombination. 

Intra-beam scattering, that occurs between the faster particles overtaking slower particles within the beam, can be minimized by reducing the velocity spread of the atomic beam from the dissociator. 

The dissociator was fine-tuned by a number of parameters: inlet gas flow, RF power, nozzle temperature and its orifice diameter. For the inlet gas flow and nozzle opening diameter, the optimal values of 0.45~$\frac{mbar\cdot l}{s}$ and 1.3~mm have been found by using the absolute beam intensity monitor that substituted the ionizer during these measurements. The atomic beam intensity, as a function of nozzle temperature and gas flow into the dissociator, shows an intensity peak at 0.45~$\frac{mbar\cdot l}{s}$ flow rate and 65~K nozzle temperature for the beam-forming system.

\begin{figure}[htbp]
\centering 
\includegraphics[width=.6\textwidth,origin=c,angle=0]{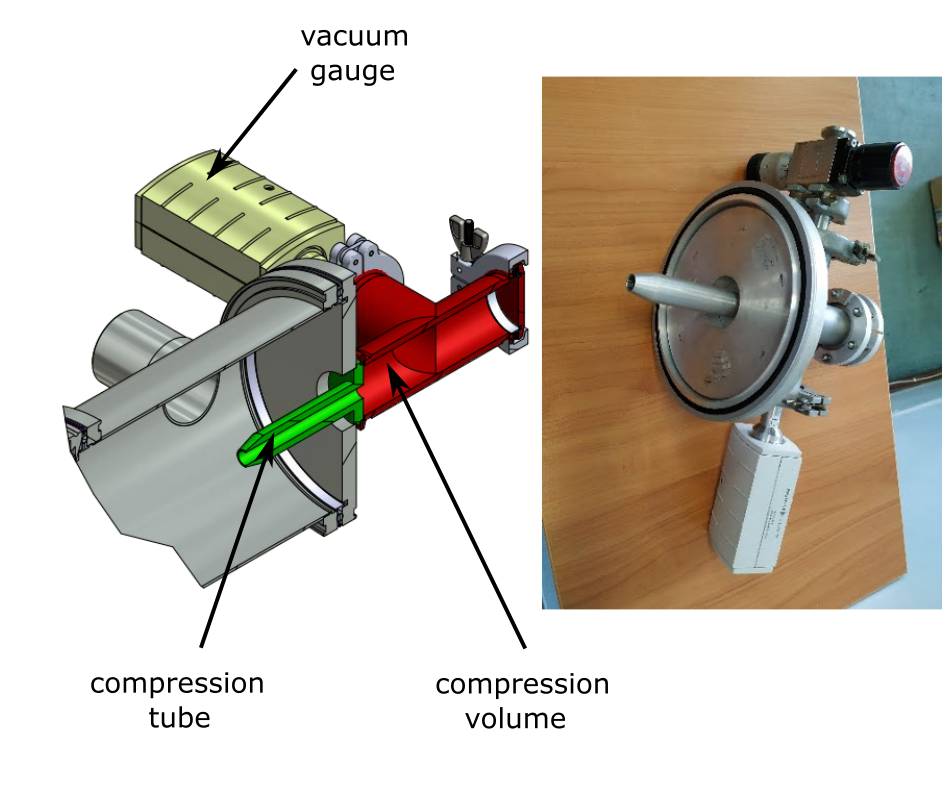}
\caption{\label{fig:h} The absolute beam intensity monitor.}
\end{figure}

The absolute beam intensity monitor shown in Figure~\ref{fig:h} is based on a compression tube that can be appropriately calibrated with a well known reference volume~\cite{j}. The beam enters a tube of 10~mm diameter and 100 mm length and the pressure rise in a compression volume is measured with a calibrated hot-cathode vacuum gauge. Since $H$ and $H_{2}$ have different gas conductance and different gauge response values, the material of the volume, stainless steel, is chosen to encourage rapid recombination of the atoms in the volume. A retractable chopper that stopped the beam was placed in front of the entrance aperture. The difference of the compression chamber pressure with chopper open and closed is proportionally dependent on the atomic beam intensity.

After the nozzle-skimmer arrangement, the deuterium atoms are subsequently fed through two sets of multipole magnets with a weak-field transition (WFT) unit placed between them. Multipole magnets forming a Stern-Gerlach device focus hyperfine states (HFS) with electron spin aligned to the magnetic field and defocus the substates with electron spin aligned in the opposite direction.

\begin{figure}[htbp]
\centering 
\includegraphics[width=.6\textwidth,origin=c,angle=0]{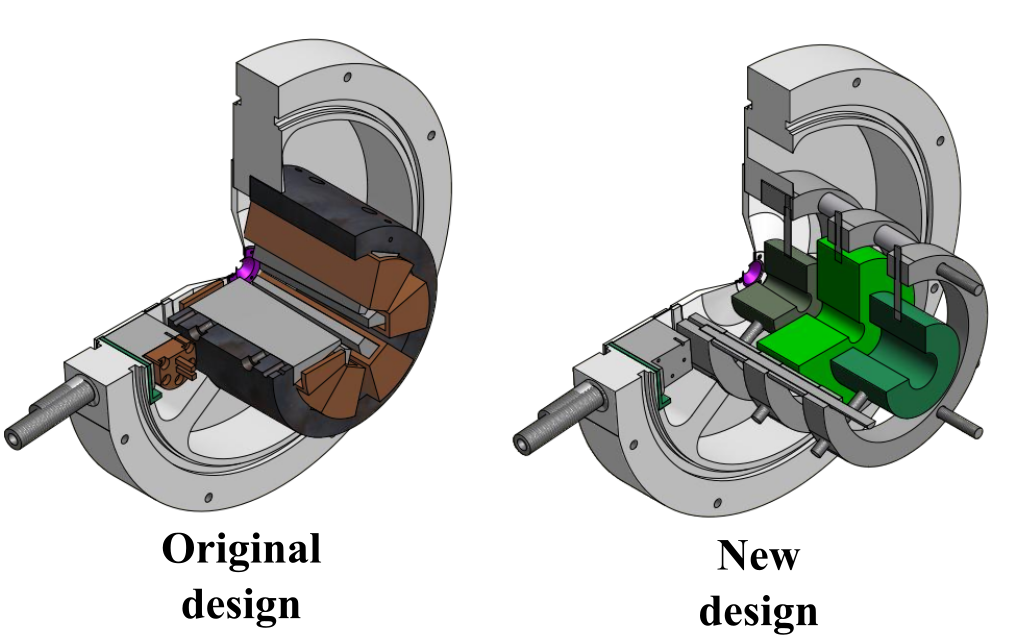}
\caption{\label{fig:l} Comparison of the original and the new design of a group of multipole magnets.}
\end{figure}

The initial design of the Stern-Gerlach device consisted of electromagnetic hexapoles with surface magnetic fields of 0.4 and 0.7~T that are induced with currents of 140 and 220~A into the corresponding coils. While placed directly in the vacuum chamber, the magnets were water-cooled with 28~bar pressure that led to large leaks of $1\cdot 10^{-5} \frac{mbar\cdot l}{s}$, which was comparable with the pressure difference originated from the atomic beam itself. Both hexapoles have been replaced with an arrangement of two groups with five multipole permanent magnets following the design proposed by Halbach~\cite{k}. The comparison with the original hexapole is shown in Figure~\ref{fig:l}. A collimator with 20~mm inner diameter is located between the groups of magnets. The geometric and magnetic parameters of the new spin-state separation magnet system are specified in Table~\ref{tab:a}. The first group represents a sextupole-sextupole-quadrupole arrangement while the other group contains only two identical sextupoles. Series of short multipoles are used to provide optimum pumping to reduce the beam losses due to scattering with the residual gas background.

\begin{table}[htbp]
\centering
\caption{\label{tab:a} New geometry of the magnets in POLIS. The magnet elements are numbered from 1-5 along the atomic beam direction:}
\smallskip
\begin{tabular}{|l|c|c|c|c|c|c|}
\hline
Element & Entrance & Length & Exit & Distance to & Number & Pole tip\\
number & i.d. [mm] & [mm] & i.d. [mm] & next element [mm] & of poles & field [T]\\

\hline
1 & 10.0 & 40.0 & 14.2 & 10.0  & 6 & 1.2 \\
2 & 15.0 & 40.0 & 15.0 & 10.0  & 6 & 1.5 \\
3 & 27.0 & 60.0 & 27.0 & 277.5 & 4 & 1.0 \\
4 & 25.0 & 62.5 & 25.0 & 27.5  & 6 & 1.5 \\
5 & 25.0 & 62.5 & 25.0 & -     & 6 & 1.5 \\
\hline
\end{tabular}
\end{table}

To minimize chromatic aberration of multipole magnets, i.e., the effect of velocity spread of particles in a beam, the magnets have been arranged in two groups separated by a drift space, and the first sextupole is tapered along the~z~axis with an inner diameter from~10~to~14.2~mm. The magnet has a~pole tip magnetic field of 1.2~T. The second sextupole and the next quadrupole have a bore of 15 and 27~mm with maximum magnetic field strengths of 1.5 and 1~T, respectively. The sextupoles of the second set have identical cylindrical shape with 25~mm inner diameter and 1.2~T magnetic field.

\begin{figure}[htbp]
\centering 
\includegraphics[width=.6\textwidth,origin=c,angle=0]{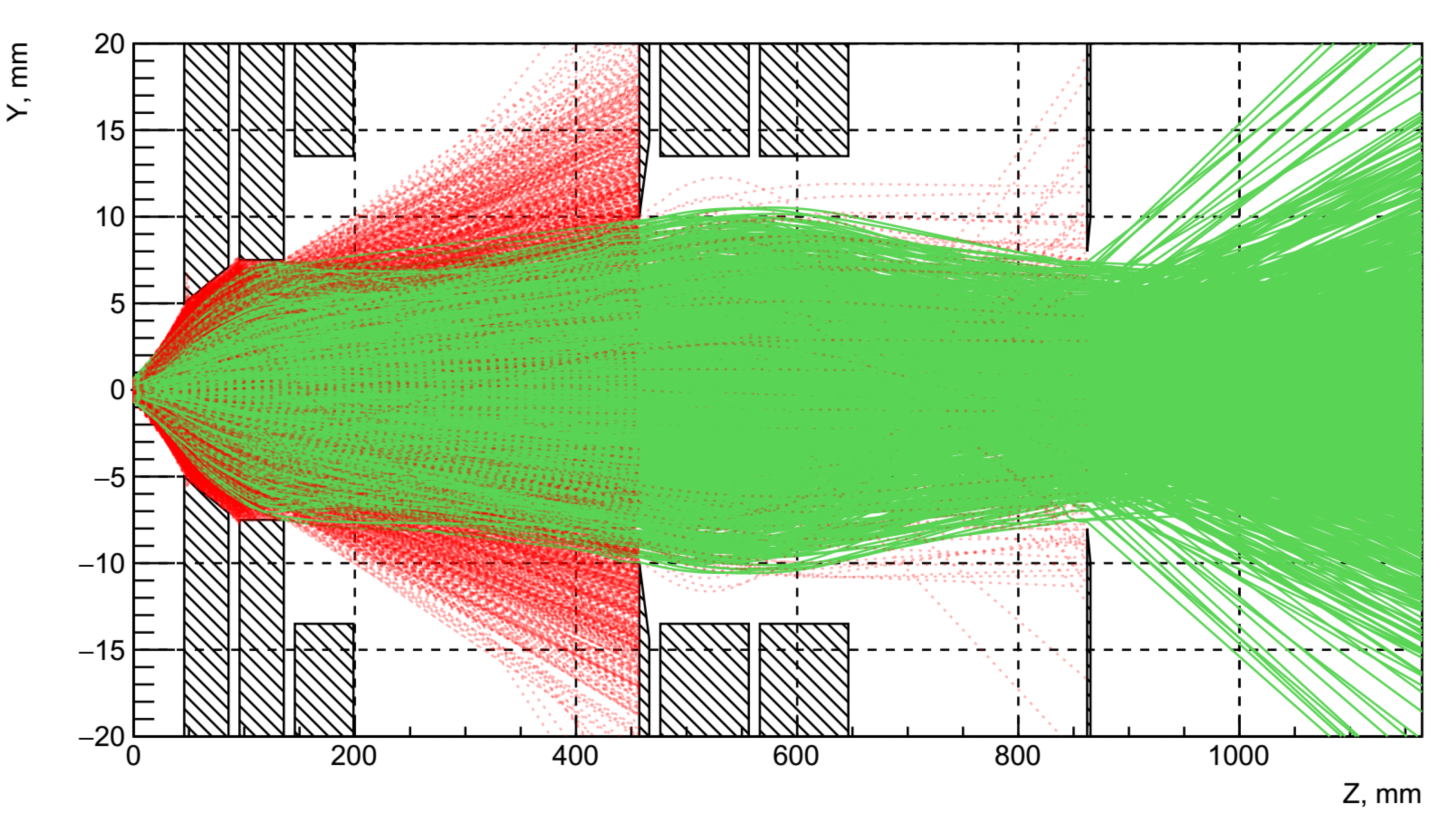}
\caption{\label{fig:d} Example of the trajectory calculation program output.}
\end{figure}

The field strengths and location of the magnets including all other apertures define the trajectories of the atomic beam through POLIS. Figure~\ref{fig:d} shows the graphic representation of the tracks mostly defined by the POLIS spin-state separation magnet system and corresponding apertures. The calculations have been performed with a Monte Carlo tracking program~\cite{l} developed in PNPI.

\begin{figure}[htbp]
\centering 
\includegraphics[width=.6\textwidth,origin=c,angle=0]{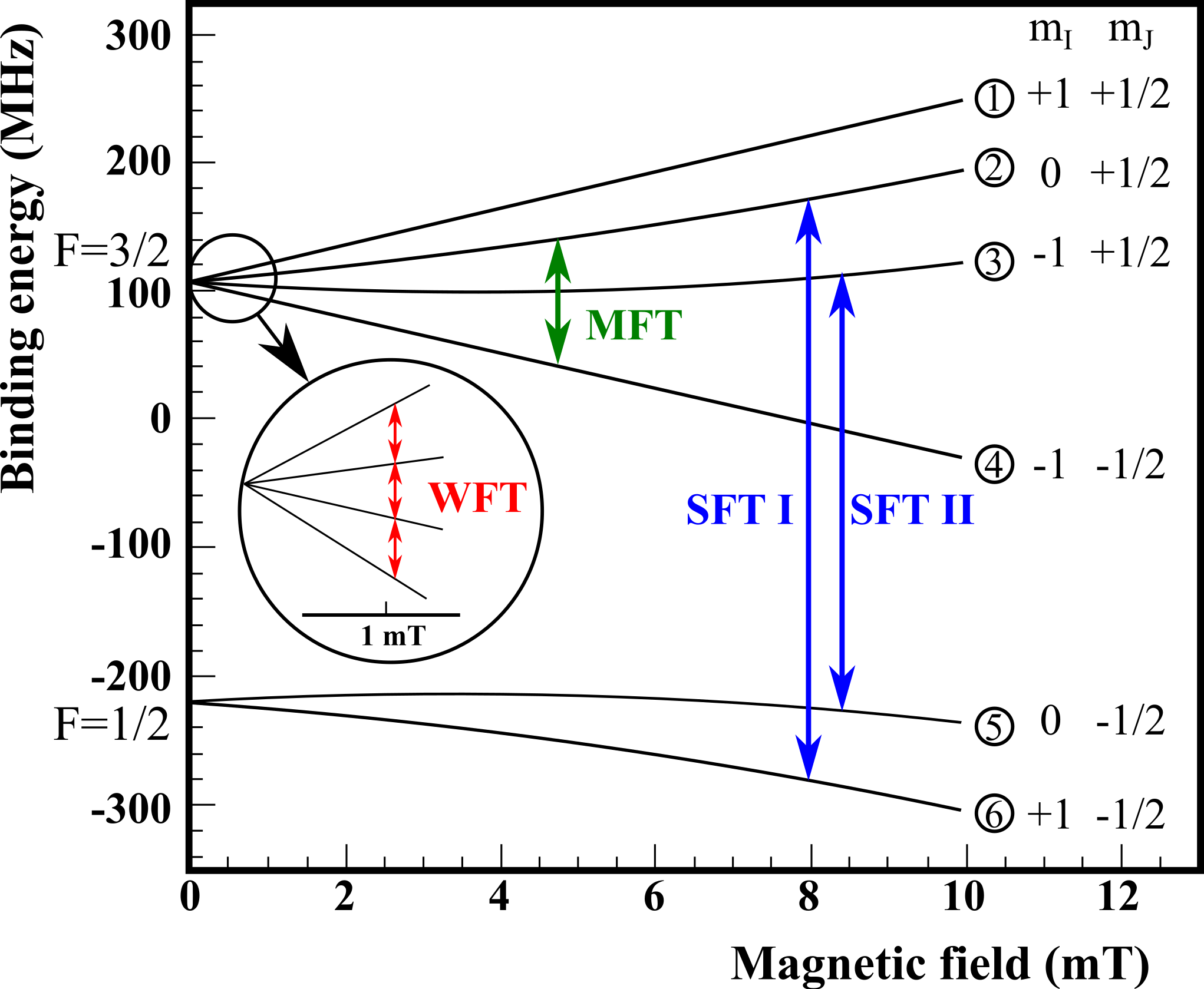}
\caption{\label{fig:f} The polarization scheme for deuterons. $m_{J}$ and $m_{I}$ denote the magnetic quantum numbers of electron and proton, respectively.}
\end{figure}

After the Stern-Gerlach part, the atoms are polarized by electron spin but nearly unpolarized in nuclear spin. To produce deuterons with certain nuclear spin orientation, the method of adiabatic passage~\cite{m} has been used. Adiabatic transitions are based on spin-state separation of the atomic electrons and transfer of the polarization to the nuclei via the hyperfine interaction. Some of the transitions are illustrated in Figure~\ref{fig:f}. This splitting occurs in the static magnetic field due to the Zeeman effect and is caused by the hyperfine interaction, i.e., by the interaction between nuclear and electron magnetic moments. After substituting the population number ($N_{+},N_{0},N_{-}$) of particles in the substates with nuclear magnetic moment $m_{I} = +1, 0$ and $-1$, respectively, into the equations~\eqref{eq:a}, vector ($p_{z}$) and tensor ($p_{zz}$) polarizations of deuteron beam can be calculated. 

\begin{subequations}\label{eq:a}
\begin{align}
\label{eq:a:1}
p_{z} &= \frac{N_{+}-N_{-}}{N_{+}+N_{-}+N_{0}}
\\
\label{eq:a:2}
p_{zz} &= \frac{N_{+}+N_{-}-2N_{0}}{N_{+}+N_{-}+N_{0}}
\end{align}
\end{subequations}

According to the strength of the static magnetic field in relation to the critical $B_{c}$ that corresponds to the HFS splitting energy, one distinguishes between weak-, medium- and strong-field transition units, simply named WFT, MFT, and SFT. POLIS utilizes three adiabatic transitions owing to the WFT and two SFT units. In the WFT unit, an RF field of 10~MHz is generated while SFT units make use of 330 and 450~MHz resonances. In these transition units, the atoms interact with a resonant RF field in a static magnetic field overlayed with a small gradient magnetic field that has a specific direction of the gradient. Under these conditions, the atoms pass the resonance adiabatically resulting in an almost complete population transfer between definite hyperfine substates~\cite{n}.

POLIS has 4 RF transition units: 2 WFT ans 2 SFT units. One of the WFT units is placed between magnets while the other WFT unit and both SFT units placed following the last set of magnets. While using a WFT induced by the second WFT unit as depicted in Figure~\ref{fig:f}, the entire population of subtate 1 is transfered into substate 4 whereas the population numbers of the substates 2 and 3 stay constant. With this adiabatic transitions, one can produce a beam of deuterons with pure negative vector polarization, i.e., $p_{z} = -\frac{2}{3}$ and $p_{zz} = 0$. To get a beam with vector polarization in opposite direction ($p_{z} = +\frac{2}{3}$ and $p_{zz} = 0$), one should use SFT units. In this case, the populations of substates 2 and 3 are transfered into substates 5 and 6, respectively. 

A beam with pure tensor polarization can be obtained by utilizing simultaneously the first WFT unit with one of the SFT units. Since the WFT unit is located between the magnets, the following set of magnets is now defocusing atoms in substate 4, thus only atoms in the states 2 and 3 stay in the beam afterwards. If now the SFT at 450 MHz is used to transfer atoms from state 2 into state 6, mostly atoms in the states 3 and 6 are found. In this case, the beam has only a positive tensor-polarization, i.e., $p_{z} = 0$ and $p_{zz} = +1$. If the SFT at 330 MHZ is used to induce transitions from substate 3 into 5, mostly atoms in the states 2 and 5 are found and the beam is negatively tensor-polarized, i.e., $p_{z} = 0$ and $p_{zz} = -2$. The adiabatic passage method provides high polarization near the theoretical maximum. However, the actual polarizations during the experiments are typically 50\%--70\% of maximum theoretical values owing to RF units perfomance and depolarization processes along the beam path.

As a result of the improvements made, we managed to achieve the atomic beam intensity at the level of $1.3\cdot{10}^{16} \frac{atoms}{s}$ measured at the center of the ionizer.

\begin{figure}[htbp]
\centering 
\includegraphics[width=.8\textwidth,origin=c,angle=0]{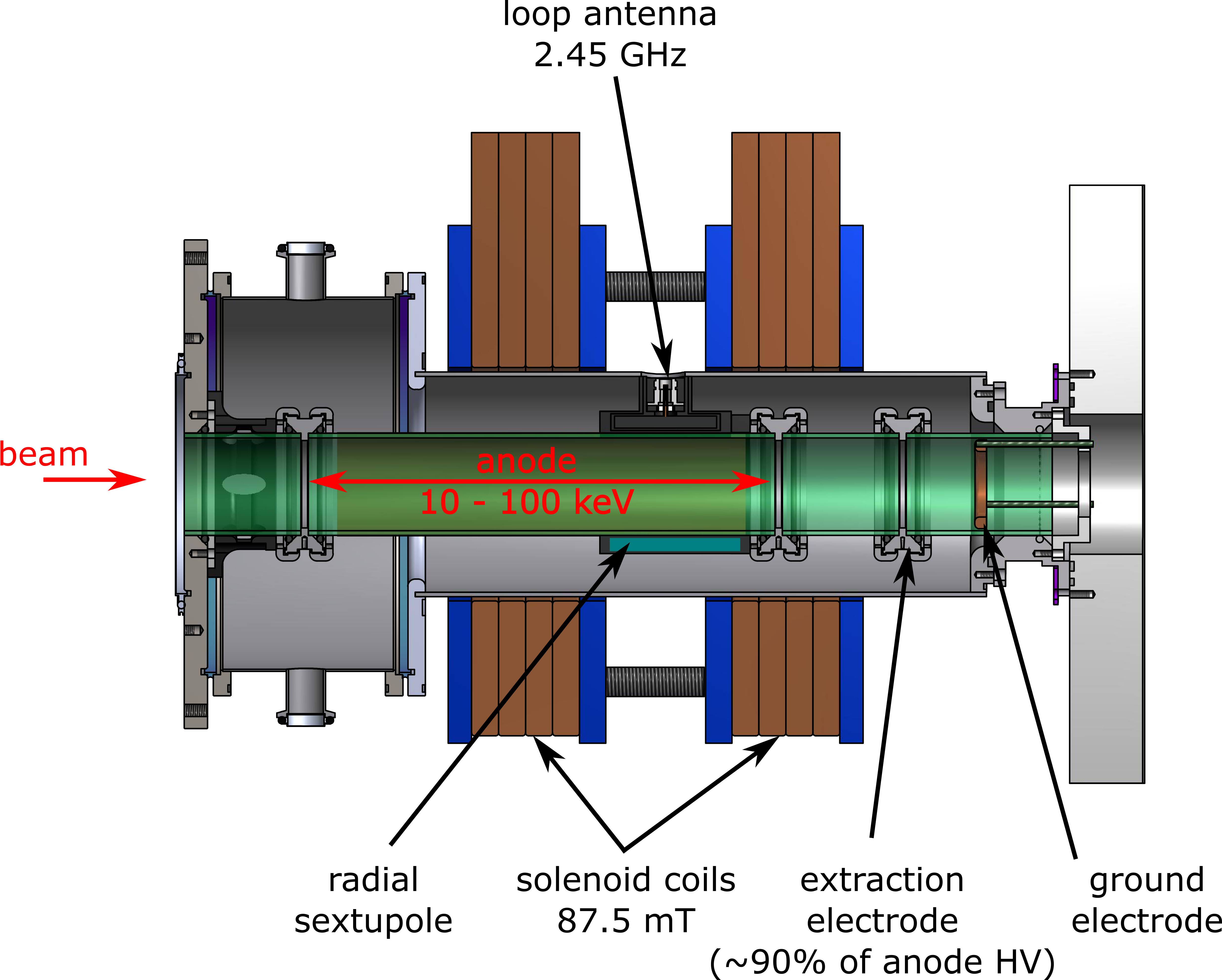}
\caption{\label{fig:j} The schematic view of the ECR ionizer.}
\end{figure}

A new ionizer based on an electron-cyclotron-resonance (ECR) has been developed. The ECR ionizer avoids the space charge problems of the intense electron beam used in other types of ionizers while having the ionization efficiency up to $6\cdot 10^{-3}$. Its schematic view and several geometric proportions are shown in Figure~\ref{fig:j} and Table~\ref{tab:b}. 

\begin{table}[htbp]
\centering
\caption{\label{tab:b} Geometry of the ECR ionizer of POLIS.}
\smallskip
\begin{tabular}{|l|c|c|c|c|}
\hline
Element & Inner 	    & Length [mm] & Distance to 	& Max \\
		& diameter [mm] & 			& next element [mm] & voltage [kV]\\
\hline
Collimator 			 & 16.0 & 10  & 40 & 0   \\
Anode 				 & 63.6 & 360 & 40 & 100 \\
Extraction electrode & 63.6 & 40  & 40 & 91  \\
Ground electrode 	 & 45.0 & 5   & -  & 0   \\
\hline
\end{tabular}
\end{table}

The ECR ionization is based on bombardment of the atoms by microwave-driven electrons. The discharge plasma inside the ionizer is strongly confined by a combination of an axial magnetic field produced by two solenoid coils and a radial magnetic field produced by a permanent hexapole magnet. In addition, microwave radiation corresponding to the ECR is coupled to the magnetic field. The fields are chosen to confine the electrons to the ionizing region and to extract the ions efficiently and without depolarization. At the solenoidal magnetic field of $\approx$87.5~mT microwave radiation of 2.45~GHz frequency is transmitted into the plasma chamber with power up to $\approx$80~W. In parallel, the magnetic field preserves the nucler polarization by decoupling the nuclear and the electron spin during the ionization process. To increase the amount of electrons and hold the plasma discharge steady, $^{4}He$ as inert gas is fed into the discharge volume at a pressure below $1\cdot 10^{-6}$~mbar. The plasma chamber consists of a 320~mm long quartz tube of 63.6~mm inner diameter allowing an intensive pumping of the chamber. 

Most parts of the ionizer, especially those connected to high voltage supplies, are installed in an envelope filled with sulfur hexafluoride ($SF_{6}$) gas used as a gaseous dielectric medium. To increase the maximum operating voltage of the ionizer up to 100~kV, the gas should be under pressure of slightly above 2~bar. Therewith, the cooling of the discharge vessel is provided by natural convection in the $SF_{6}$ atmosphere.

One of the basic requirements in order to minimize the background scattering of the beam particles is a powerful differential pumping system. In almost every chamber, a high-performance turbopump is installed. It provides a pressure inside the ionizer in the order of $10^{-7}$~mbar in the presence of the atomic hydrogen beam with a gas flow into the dissociator of about 0.45~$\frac{mbar\cdot l}{s}$.

A new microcontroller-based control system with various interlocks has been developed and commissioned to improve the reliability of the setup with smooth online control and monitoring.

POLIS is now producing a deuteron beam current of $\approx 15\:\mu$A on the target. These improvements of POLIS led to a stable ion beam production with a continuous performance over a period of about two weeks. 

\subsection{The polarized atomic beam source (ABS) for the jet target}
\label{sec:ABS}

The University of Ferrara contributed the polarized atomic beam source that was previously used at the Indiana University Cyclotron Facility (IUCF) and known as the Wisconsin ABS~\cite{o}. It is able to deliver a beam of deuterium atoms of requested nuclear vector and tensor polarization and energies of about 0.01~eV. The initial intensity of the beam $4\cdot 10^{16}\:\frac{atoms}{s}$ is sufficient for the experiment to investigate the fusion reactions with low cross-section. 

\begin{figure}[htbp]
\centering 
\includegraphics[width=.6\textwidth,origin=c,angle=0]{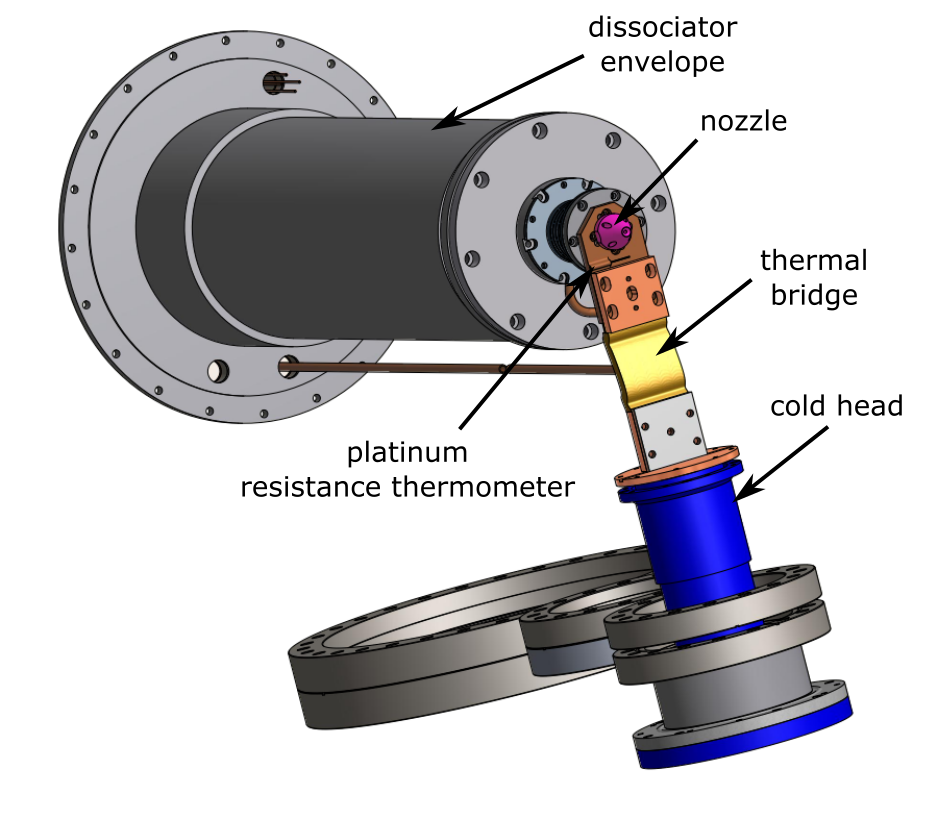}
\caption{\label{fig:q} The ABS dissociator cooling system.}
\end{figure}

Several changes to the original dissociator have been made by the PNPI group to improve reliability and performance of the ABS. Initially, the nozzle had been cooled down to 84~K directly by liquid nitrogen streamed through a tube soldered around the nozzle~\cite{o}. To reduce maintenance, a cold-head refrigeration system has been assembled to cool the nozzle by a thermal bridge as illustrated in Figure~\ref{fig:q}. The temperature is monitored by using a platinum resistance thermometer PT-100 attached near the nozzle. 

\begin{figure}[htbp]
\centering 
\includegraphics[width=.8\textwidth,origin=c,angle=0]{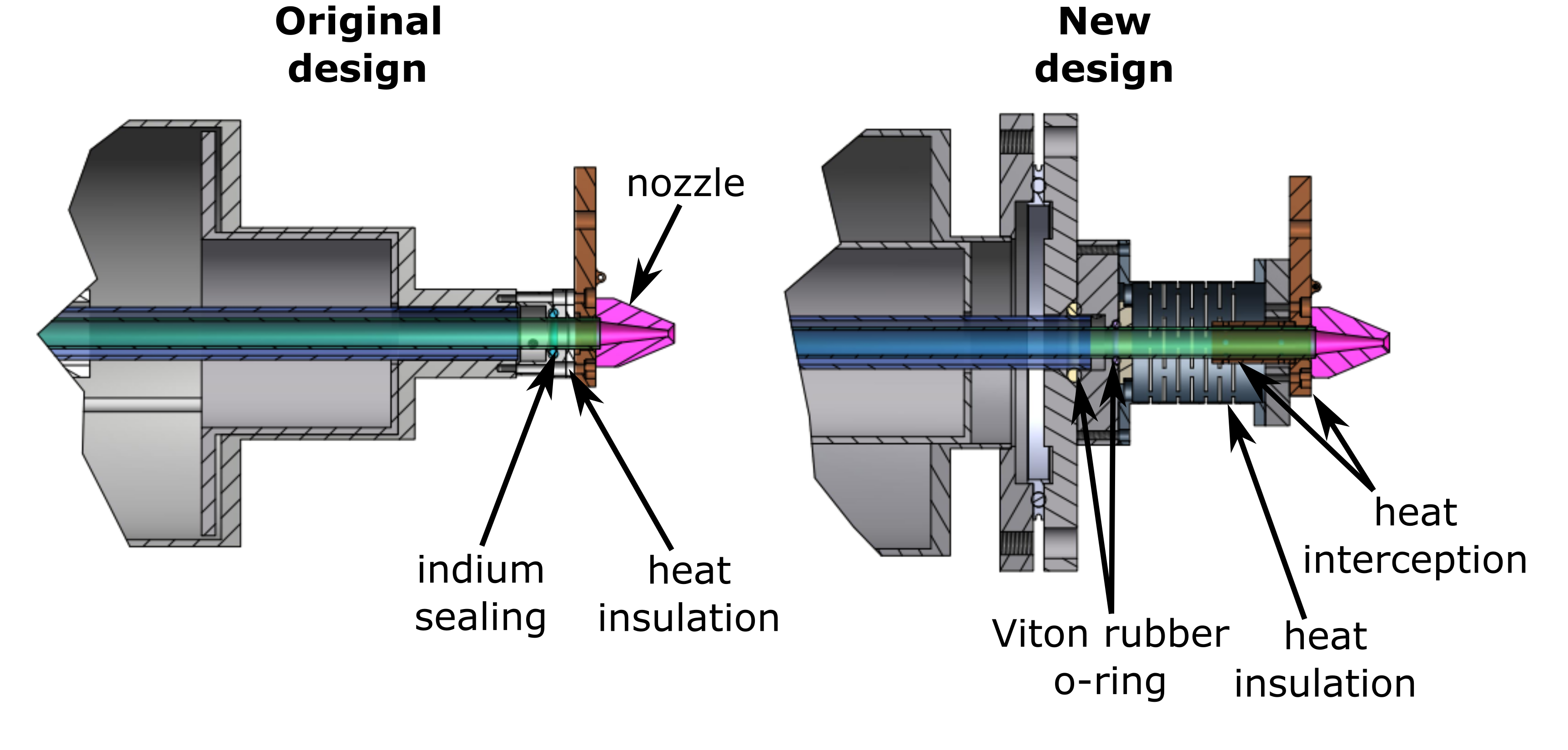}
\caption{\label{fig:r} Comparison of the original and new design of the dissociator nozzle.}
\end{figure}

To facilitate disassembly of the dissociator, a disposable indium sealing has been replaced with Viton rubber O-rings while increasing the distance to the nozzle as shown in Figure~\ref{fig:r} to prevent the rubber from overcooling. The dissociator tube is cooled by water. With the new refrigeration system, it becomes possible to cool the nozzle down to 40~K and have a better possibility to optimize the nozzle temperature. The RF generator of the dissociator has been replaced with a new one. A special matching circuit, equipped with a reflectometer, has been developed to match the new generator with the dissociator load to reduce the reflected power of the initial value of 300~W to only 3~W.

The ABS provides an atomic beam based on the same principles as POLIS. Parameters of the electron spin-state separation magnet system are specified in Table~\ref{tab:c}. The ABS makes use of two groups of 3 sextupoles.

\begin{table}[htbp]
\centering
\caption{\label{tab:c} Magnet geometry of the ABS. The magnet elements are numbered from 1-6 along the atomic beam direction.}
\smallskip
\begin{tabular}{|l|c|c|c|c|c|c|}
\hline
Element & Entrance & Length & Exit & Distance to & Number & Pole tip\\
number & i.d. [mm] & [mm] & i.d. [mm] & next element [mm] & of poles & field [T]\\
\hline
1 & 10.5 & 25.0 & 10.5 & 10.0 	& 6 & 1.36 \\
2 & 12.0 & 50.0 & 16.5 & 10.0 	& 6 & 1.36 \\
3 & 16.6 & 60.0 & 22.0 & 12.0	& 6 & 1.36 \\
4 & 25.0 & 35.0 & 25.0 & 360.0	& 6 & 1.50 \\
5 & 25.0 & 62.5 & 25.0 & 27.5 	& 6 & 1.50 \\
6 & 25.0 & 62.5 & 25.0 & - 		& 6 & 1.50 \\
\hline
\end{tabular}
\end{table}

Contrary to POLIS, the ABS utilizes two identical MFT units that are able to operate under WFT and MFT conditions. In addition, an MFT unit allows to select different transitions by changing the static magnetic field by a small amount. One of the units is placed between two sets of sextupole magnets while the other is located right after the last sextupole. 

As an example, to obtain a beam with pure positive tensor polarization, i.e., $p_{z} = 0$ and $p_{zz}=+1$, the first MFT unit induces a spin state transition 2 into 4 in notations of Figure~\ref{fig:f}. After the unit, the second permanent sextupole magnet group focuses spin states~1 and 3, but defocuses spin state~4 atoms leading to polarized deuterium atoms in spin states~1 and 3 with equal populations. Other variations of the polarization can be obtained with a similar procedure by using both MFT units.

The beam-forming system in general is almost the same as at POLIS but with one significant difference: The ABS has two apertures between the nozzle and the magnet system, a skimmer and a collimator, while POLIS has only a skimmer. Dividing the space between the nozzle and the sextupole entrance into two short sections significantly reduces the loss by residual gas scattering. 

As POLIS, the ABS has a multistage differential pumping system designed to avoid intensity losses due to background gas scattering. 

The automatic control system for monitoring and handling of the source has been implemented and combined with the PolFusion control system. 

\section{Polarimetry}
\label{sec:polarimetry}

The use of polarized atomic and ion beams requires the knowledge of their nuclear polarization. It is planned to use two polarimeters in the experiment: a Lamb-shift polarimeter and a nuclear-reaction polarimeter. 

\subsection{Lamb-shift polarimeter (LSP)}
\label{sec:LSP}

In preparation for the experimental runs, the LSP will be used to tune both, the ABS and POLIS. During the production run, it is positioned after the $4 \pi$ detector system in conjunction with a Glavish type ionizer~\cite{p} for online-monitoring of the ABS polarization. An LSP~\cite{q} allows a fast determination and precise measurements of the vector and tensor polarization of hydrogen isotopes.

\begin{figure}[htbp]
\centering 
\includegraphics[width=.6\textwidth,origin=c,angle=0]{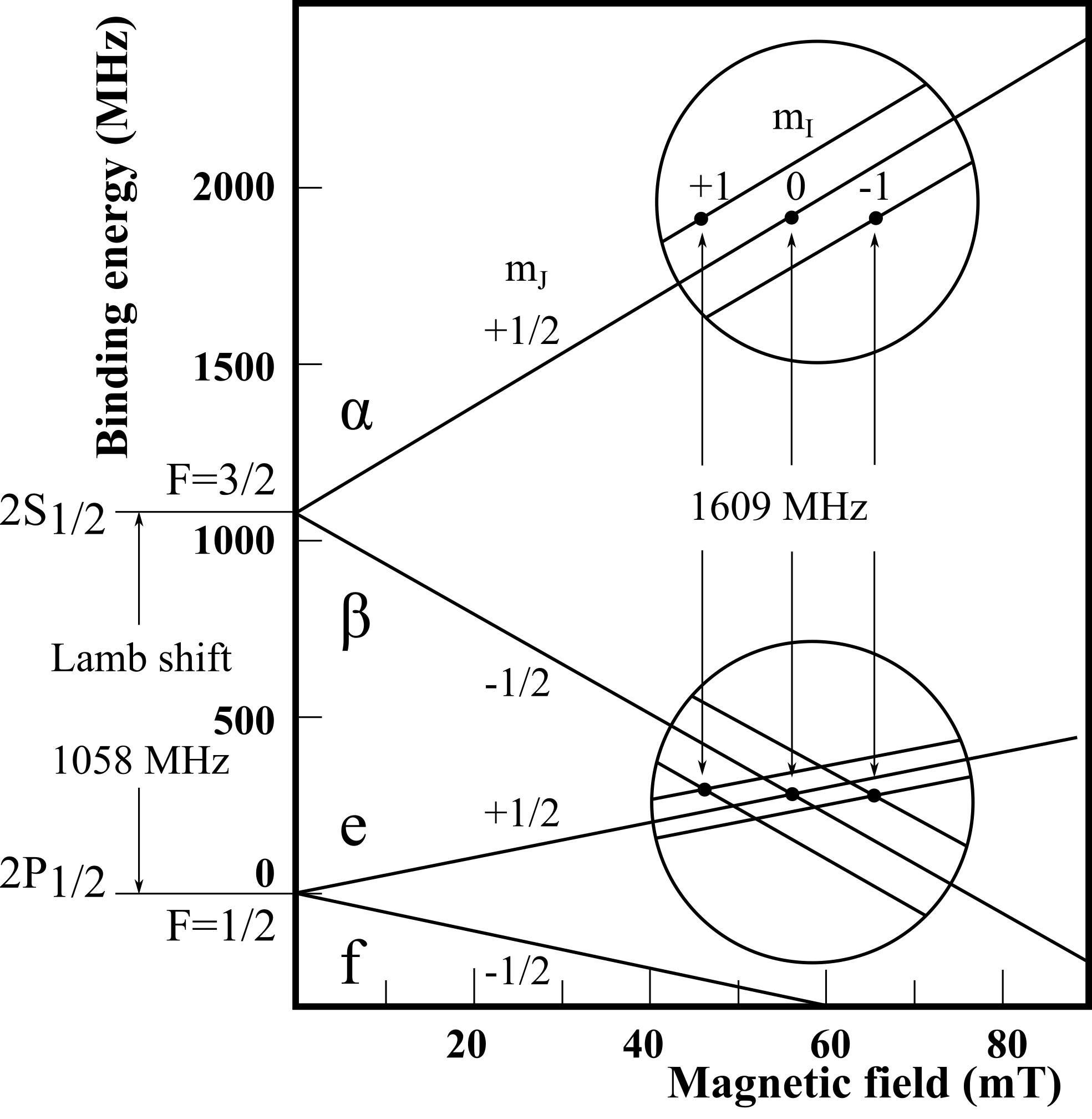}
\caption{\label{fig:a} The magnetic field dependency of the binding energies of the hyperfine substates for deuterium atoms.}
\end{figure}

The LSP utilizes a small difference in energy between the $2S_{\frac{1}{2}}$ and $2P_{\frac{1}{2}}$ levels of the hydrogen atoms called the Lamb-shift. The existence of the shift allows mixing of the metastable $2S_{\frac{1}{2}}$ hyperfine states with the short-living $2P_{\frac{1}{2}}$ states in external magnetic and electric fields to manipulate the lifetime of metastable atoms in a specific quantum state. Figure~\ref{fig:a} shows the Breit-Rabi diagram of the deuterium atom in $2S_{\frac{1}{2}}$ and $2P_{\frac{1}{2}}$ states. 

Atoms with an electron in a spin-up (spin-down) state are labeled with $\alpha$ and $e$ ($\beta$ and $f$). At a magnetic field of 56.5, 57.5 and 58.5~mT, $\beta$ and $e$ substates crosses each other, i.e., the binding energies are degenerated. At the same time, they have the same energy separation (1609~MHz) from the corresponding $\alpha$-state. The LSP utilizes a three-level interaction between $\alpha$, $\beta$ and $e$~states of~the deuterium atoms. In detail, the parity violation of the electric field (Stark effect) couples the $\beta$ to the $e$ state and the RF field couples the $\alpha$ states to the corresponding $e$ states. By that, the lifetime of the $\alpha$ and the $\beta$ states can be manipulated with the strength of the electric and the RF field. Thus, at a special magnetic field metastable atoms in a single $\alpha$ substate can be captured in an oscillation with the correspondig $\beta$ state while all other metastable atoms will decay into the ground state. The~magnetic field strengths mentioned above correspond to~the~specific hyperfine states $\alpha _{1}\:|m_{J}= \frac{1}{2}, m_{I}= 1 \rangle$, $\alpha _{2} \:|m_{J}= \frac{1}{2}, m_{I}= 0 \rangle$ and $\alpha _{3} \:|m_{J}= \frac{1}{2}, m_{I}= -1 \rangle$.

\begin{figure}[htbp]
\centering 
\includegraphics[width=.8\textwidth,origin=c,angle=0]{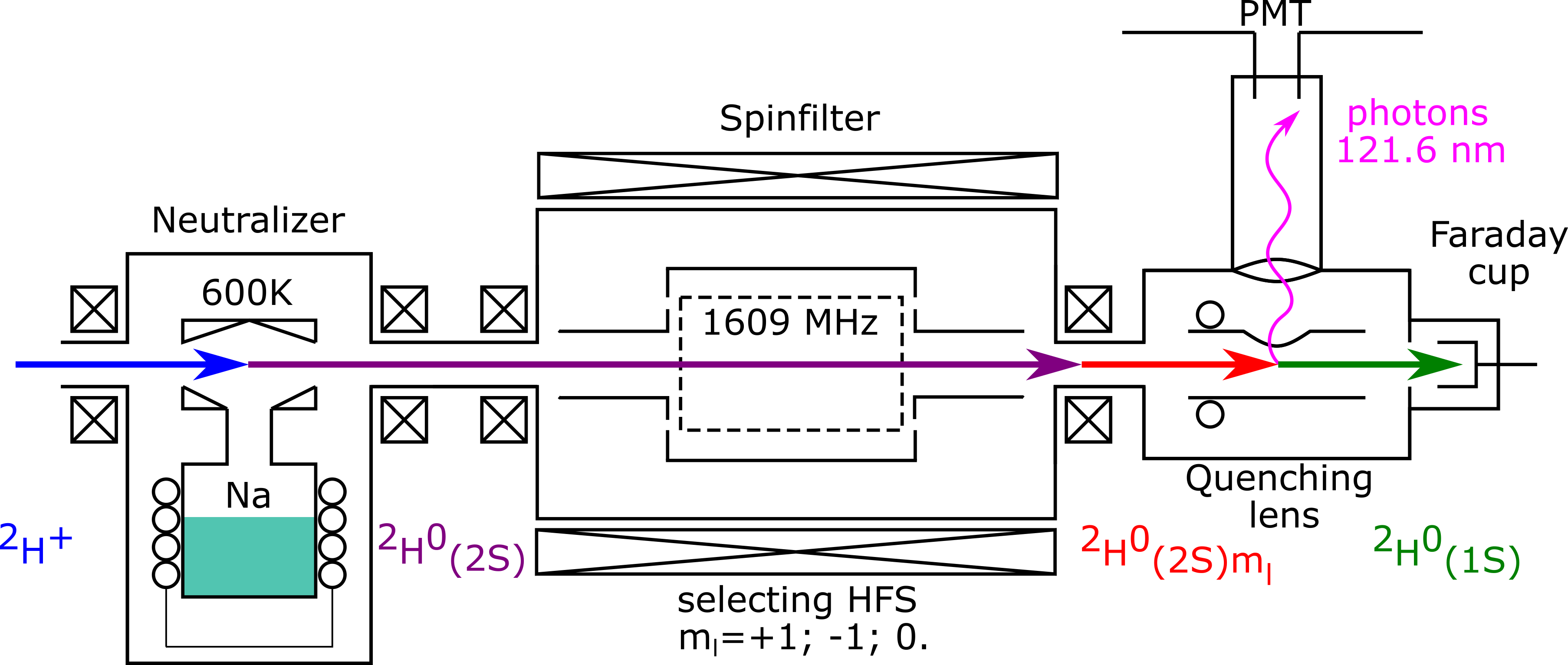}
\caption{\label{fig:b} The schematic representation of the LSP.}
\end{figure}

The schematic setup of the LSP is shown in~Figure~\ref{fig:b}. It includes three main parts: a neutralizer, a spinfilter, and a quench chamber. 

The neutralizer is a gas target of sodium vapor which is used to produce metastable deuterium in the first excited state $2S_{\frac{1}{2}}$ through the charge exchange reaction $^{2}H^{+}+Na\rightarrow \:^{2}H(2S)+Na^{+}$. A longitudinal magnetic field of about 50 mT is necessary to minimize the depolarization.

The metastable atomic beam travels through the spinfilter~\cite{r}. At a certain solenoidal magnetic field strength, only metastable atoms in a definite hyperfine substate can pass the spinfilter while other states are quenched into the ground state. The amount of metastable atoms leaving the spinfilter at a given magnetic field is proportional to the amount of deuterons entering the neutralizer with the corresponding nuclear spin.

In the quench chamber, the metastable atoms, which survived in the spinfilter, are quenched into the ground state by a strong electric field of an electrostatic lens due to the Stark effect. The emitted $Ly-\alpha$ photons are registered with a dedicated photomultiplier as a function of the magnetic field strength in the spinfilter. 

Comparing the amount of measured Lyman-alpha photons by using the formulae~\eqref{eq:a} delivers vector- and tensor-polarization with a precision better than 1\% within a few seconds.

\subsection{Nuclear reaction polarimeter (NRP)}
\label{sec:NRP}

Due to the necessity of measuring the polarization of the deuteron beam with higher energies than the LSP can handle, an NRP~\cite{s} is used for long-term measurements and monitoring of POLIS. The NRP is based on the DD reaction $^{2}H(d,p)^{3}H$ with use of an unpolarized target.

POLIS delivers a longitudinally polarized beam. To allow the determination of both, the vector and tensor polarization components, the polarimeter requests transverse-polarized ions. For this purpose, an electrostatic deflector is used and placed between POLIS and the NRP.

Inside the scattering chamber of the NRP, a titanium-foil target is placed in the center surrounded by nine semiconductor detectors (type Hamamatsu S3508-09 with $10 \times 10$~mm$^2$ active area) at a distance of 100~mm from the target. These PIN diodes can be rearranged at polar angles between $10^{\circ}$ and $170^{\circ}$ relative to the direction of the incoming beam with two orthogonal ring-shaped supports.

To minimize the energy loss on the target and energy dispersion of the reaction products, the titanium foil has to be as thin as possible, e.g., 1~$\mu$m. The target can be prepared by bombarding the foil with sufficient beam deuterons or heating the foil in vacuum for outgassing with subsequent cooling down in deuterium atmosphere. Besides, the target can be made by vacuum evaporation of deuterated parapolyphenyl, $(C_{6}D_{4})_n$, or deuterated polyethylene, $(CD_{2})_n$, onto a~thin titanium foil or a foil of another appropriate material.

To determine the polarization components by using the~NRP, one requires to use the~vector and tensor asymmetries, obtained previously by other researchers~\cite{d,e,f}, in the count rates of the protons and tritons between the certain detectors at different azimuthal angles. 

\section{Detector system}
\label{sec:detector system}

One of the crucial tasks of the experiment is the detection of the fusion-reaction
ejectiles, i.e., protons, tritons, and $^{3}He^{2+}$ ions. Neutrons can not be registed with this experimental setup. Due to the positive Q-value of these fusion reactions of about 4~MeV, all reaction products have energies essentially higher than the initial deuterons from POLIS. The energy of the reaction products is determined from kinematics of a binary interaction and corresponds to 0.8~MeV for the helion, 1.0~MeV -- triton and 3.0~MeV -- proton. Background signals from scattered electrons and deuterons can be discriminated against the ejectile spectra. 

As has been mentioned above, both beams are crossing in the center of the $4 \pi$ detector system to measure the angular distributions of the reaction products. The $4 \pi$ detector consists of 576 Hamamatsu S3508-09 silicon PIN photodiodes  arranged in a cubic structure and attached on its inner surface. The thickness of the PIN diodes sensitive layer is 0.3~mm. The center of each plane has a square window to improve the conditions of vacuum pumping and also for conducting beams through front and left sides of the cube. The~PIN diodes placement allows the effective coverage of 51\%. The~$4 \pi$ detector system can measure fusion products energy in the~range of 200~keV to 4~MeV. 

A special readout electronic module ASF48 was designed and manufactured in PNPI for the data acquisition from the $4 \pi$ detector system. Each module contains 48 charge-sensitive channels. The module provides continuous and simultaneous measurement of all 48 channels with six eight-channel pipeline ADCs. The sampling frequency of the ADC is 65~MHz. The signal in every channel is registered automatically in case of exceeding the amplitude trigger threshold and is sent to the MIDAS data acquisition system for on-line monitoring and data storage for off-line analysis. The time of the signal is also recorded by the trigger actuation with 10~ns accuracy. 

\section{Test run}
\label{sec:test}

In 2019, an unpolarized experiment with an ion beam from POLIS and heavy water vapor as a target has been carried out in order to check the performance of POLIS at the lowest energy at 10~keV, the $4 \pi$ detector system readout electronics, and the cosmic background sensitivity. 

\begin{figure}[htbp]
\centering 
\includegraphics[width=.6\textwidth,origin=c,angle=0]{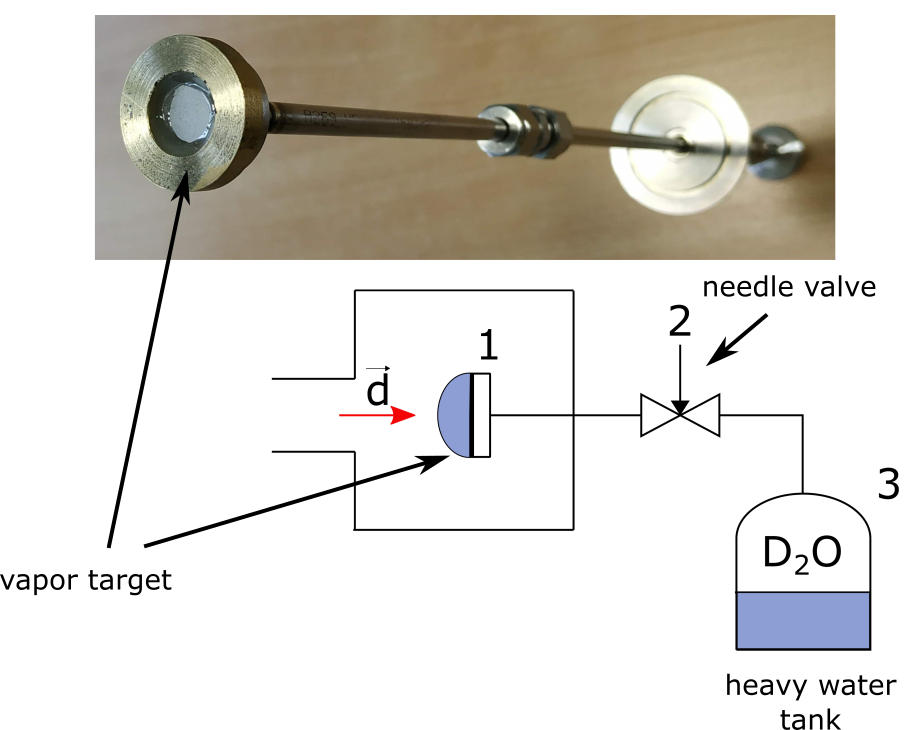}
\caption{\label{fig:o} Vapor target and its schematic representation.}
\end{figure}

The heavy water vapor target, presented in Figure~\ref{fig:o}, consists of a porous stainless steal foil (1), the fine adjustment valve (2), and the heavy water volume (3). After passing a needle valve, heavy water instantly evaporates due to the large pressure difference. As a result, the $D_{2}O$ gas passes through the porous stainless steel. The flow of heavy water vapor, which defines the density of the target, is controlled by adjusting the valve (2) in accordance with the pressure changes in the scattering chamber. After some tests, the resulting flow has been fine-tuned to $\approx {10}^{18} \frac{atoms}{s}$. 

The vapor target has been installed in the $4 \pi$ detector system with only 25 silicon detectors in total attached on front, bottom and top planes. The target was exposed to the ion beam of 15~$\mu$A and 10~keV energy. 

\begin{figure}[htbp]
\centering 
\includegraphics[width=.6\textwidth,origin=c,angle=0]{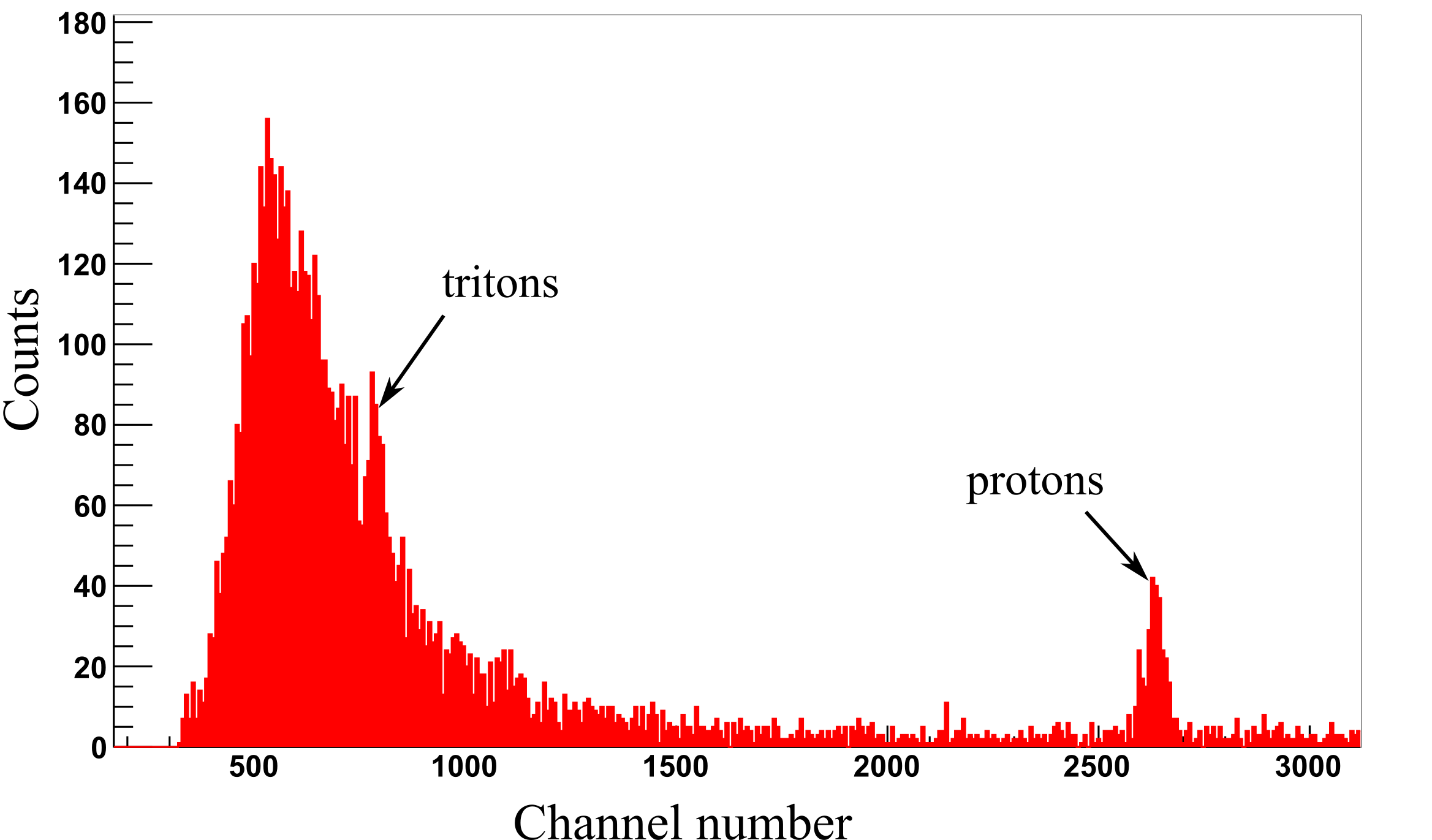}
\caption{\label{fig:n} Amplitude spectrum of unpolarized DD fusion reactions at 10 keV with cosmic background.}
\end{figure}

The charge-particle spectrum at 10~keV is shown in Figure~\ref{fig:n} with channel 2600 corresponding to the energy of about 3~MeV. The amplitude spectrum shows two distinct peaks of the deuterium fusion products for tritons and protons. However, the third peak, 0.82~MeV for $^{3}He^{2+}$, merges with the cosmic background. To distinguish these rare events in future runs, the background will be discriminated by scintillator counters placed above the scattering chamber and connected in anti-coincidence with the detectors.

Successful registration of the DD fusion reaction products on top of the cosmic background confirms the possibility of 10~keV reaction investigation and the sensitivity of the readout electronics. The observed count rate for the charged particles from the DD reactions using the vapor target was about 3 events per hour.

\section{Summary}
\label{sec:sum}

The PolFusion experiment aims to measure (among other quantities) angular distributions of the differential cross-sections with different polarization of two crossing beams with a center-of-mass energy from 10 to~100~keV. 

POLIS has undergone the following changes: improvements of the beam-forming system and RF circuit, replacement of electromagnetic hexapoles with permanent magnets, and development of a brand new ECR ionizer capable to operate at 100 kV. For an inlet gas flow of 0.45~$\frac{mbar\cdot l}{s}$ and 1.3~mm nozzle opening diameter, an atomic beam intensity of $1.3\cdot{10}^{16} \frac{atoms}{s}$ has been measured at the center of the ionizer. Due to the optimized ionizer with a solenoidal field of $\approx$87.5~mT, microwave radiation of 2.45~GHz frequency and power of $\approx$80~W, a deuteron beam current of $\approx 15\:\mu$A at 10~keV energy has been obtained. There are still several options to increase the current further.

For the ABS to produce the jet target, a cold-head refrigeration system has replaced the liquid nitrogen cooling of the nozzle. In addition, the dissociator has been modified with Viton rubber O-rings instead of the former indium sealing. In addition, the reflected power of the RF dissociator circuit has been reduced from 300 W to 3 W.

The polarimetry system of the experiment has been described including a Lamb-shift polarimeter and a nuclear-reaction polarimeter based on the DD reaction $^{2}H(d,p)^{3}H$. A sodium cell is proposed to be used as a neutralizer in the LSP.

An unpolarized experiment with an ion beam of 10~keV energy and heavy water vapor as a target has been carried out. The charged-particle spectrum at 10~keV demonstrates the sensitivity and performance of the setup.

Our future activities will concentrate on the polarization values of the deuteron/deuterium beams of POLIS and the jet target by monitoring these values with the LSP, including the new sodium neutralizer. In addition, the scintillator counters for background suppression in the spectra will be installed.


\end{document}